\documentstyle[seceq]{ptptex}

\title{Towards a Topological Formulation of Fundamental Interactions}

\author{Marco {\sc Spaans}\footnote{Hubble Fellow, E-mail:
mspaans@cfa.harvard.edu}}

\inst{Harvard-Smithsonian Center for Astrophysics, 60 Garden
Street, MS 51, Cambridge, MA 02138, USA}

\recdate{
\today
}

\abst{
A thought experiment is formulated to unify quantum mechanics
and general relativity in a topological manner.
An analysis of the interactions in Nature is then presented.
The universal ground state of the constructed theory derives from the cyclic
properties ($S^1$ homotopy) of the topological
manifold $Q=2T^3\oplus 3S^1\times S^2$ which has 23 intrinsic degrees of
freedom, discrete $Z_3$ and $Z_2\times Z_3$ internal groups, an $SU(5)$ or
$SO(10)$ gauge group, and leads to an anomalous $U(1)$ symmetry on a lattice.
These properties can in principle reproduce the standard model
with a stable proton. The general equation
of motion for the unified theory is derived up to
the Planck energy and leads to a Higgs field with possible inflation. The
thermodynamic properties of $Q$ are discussed and yield a consistent
amplitude for the cosmic microwave background fluctuations. The manifold $Q$
possesses internal energy scales which are independent of the field theory
defined on it, but which constrain the predicted mass hierarchy of such
theories. In particular the electron and its neutrino are identified as
particle ground states and their masses are predicted. The mass of the electron
agrees very well with observations. A heuristic argument for the occurrence and
magnitude of CP violation is given. Future extensions of the presented
framework are discussed.
}

\begin{document}

\maketitle

\section{Introduction}

One of the outstanding questions in quantum cosmology and particle physics
is the unification of gravity with the electro-weak and strong interactions.
Much effort has been devoted in the past years to formulate a purely
geometrical and topological theory for both types of interactions[1,2].
Probably best known are the theories involving super-gravity and
superstrings[1]. These theories have been shown recently to be unified in
M-theory, although the precise formulation of the latter is not known yet.
Three outstanding problems in these approaches are compactification down to
four dimensions, the existence of a unique vacuum state, and most importantly
the formulation of a guiding Physical Principle to lead the mathematical
construction of the theory. This work aims at extending, in a {\it physical}
manner, the topological dynamics approach presented in [3] (paperI from here
on), which derives from the {\it mathematical} consistency of multiply
connected space-times when they required to be Lorentz invariant and to
support quantum mechanical superposition.

The relevant results of paperI are as follows.
The properties of space-time topology are governed by homotopically
inequivalent loops in the {\it prime} manifolds $S^3$,
$S^1\times S^2$, and $T^3=S^1\times S^1\times S^1$. The three-torus and the
handle are nuclear, i.e.\ they bound a Lorentz four-manifold with $SL(2;C)$
spin structure. This is the only set of primes which assures Lorentz
invariance and the superposition principle as formulated by the Feynman path
integral for all times, {\it if} space-time on the Planck scale can be
described by a (3+1)-dimensional topological manifold.
The dynamics of the mathematical theory are
determined by the loop creation ($T^\dagger$) and loop annihilation ($T$)
operators, which obey $[T,T^\dagger ]=1$.
Their actions on a manifold $M$ are $T^\dagger M=S^1\times M$ and
$TM=nM'$, with $n$ the number of loops in $M$ and $M'$ the manifold $M$ with
a loop shrunk to a point. On the Planck scale,
space-time has the structure of a
{\it lattice} of Planck size three-tori $L(T^3)$ with 4 homotopically
inequivalent paths joining where the $T^3$ are connected through
three-ball surgery. The existence of this four-fold symmetry
leads to $SO(n)$ and $SU(n)$ gauge groups as well as
the numerical factor $1/4$ in the expression for black hole entropy
(but not the quantum states).
The number of degrees of freedom of the prime manifolds,
referred to as the prime quanta, under the action of
$O=TT^\dagger +T^\dagger T$, are 1, 3, and 7 for the three-sphere,
the handle manifold, and the three-torus, respectively.
During the Planck epoch, $S^1\times S^2$ handles (mini black holes) can
attach themselves to the $T^3$ lattice which leads to additional field
interactions between the degrees of freedom of the prime manifolds (to be
studied in this
work). The $S^1\times S^2$ prime manifolds are referred to as charges on the
$T^3$ lattice from hereon. Finally, it was found that the cosmological
constant $\Lambda$, viewed as the spontaneous creation of mini black
holes from the vacuum, is very small and proportional to the number of
(macroscopic) black holes at the current epoch.
Since mini black holes evaporate as the universe
expands, this provides a natural decrease of the late time cosmological
constant, but implies that any macroscopic black hole increases in mass $M$
proportional to $\Lambda M^3$. As mentioned, these results
follow from the required {\it mathematical} consistency of a theory
based on fundamental prime manifolds. They constitute a generalization of
Mach's principle as interpreted by Einstein in which inertial forces
are a consequence of the global geometry {\it as well as}
topology of the universe. The thought experiment given below will provide a
{\it physical} basis for these results.

This paper is therefore self-contained, and is organized as follows. Section 2
presents a thought experiment which relates forces (including gravity) to
the topology of space-time.
Section 3 discusses the general properties of a Quantum-gravitational Grand
Unified Theory (QGUT) based on the charged $T^3$
lattice and presents the derivation of the fundamental equation of motion for
the wave function. The equation of motion constitutes the main result of this
work. Section 4 discusses the symmetry groups of $Q$ and their relation to the
standard model. Section 5 contains the conclusions and discussion.

\section{Space-time Topology and the Nature of Forces}

The existence of the prime manifold $T^3$ in a lattice allows the mathematical
implementation of the superposition principle as in the Feynman path integral,
while the nuclearity of the three-torus and the handle guarantees Lorentz
invariance of the Planck scale topological manifold (paperI). Nevertheless,
one needs to formulate an independent Physical Principle which {\it leads} to
$T^3$ and $S^1\times S^2$ so that both superposition and Lorentz invariance
are derived concepts. To find this Physical Principle, which must accommodate
general relativity, the following thought experiment is performed.

\subsection{Thought Experiment}

Imagine an observer with a measuring rod of accuracy $\ell$.
This same observer is located in a zero gravity
environment to witness the motion of an object
and to record its position. He can conclude from measurements
of its trajectory that a force acts on the object. As he performs
similar experiments for smaller and smaller scales he notices that a smooth
description of the observational data
is limited by the accuracy of his measuring rod. The trajectories of various
objects still appear to be continuous but derivatives are very uncertain.
Any set of measurements therefore leaves room
for the conclusion on his part that something prevents the particle
trajectories from becoming identical on scales $\le\ell$, although they very
well could be identical within the error bars.
In the former case, the physical obstruction responsible
is hidden by his inability to measure accurately. As the
observer repeats his experiments for all orientations in three-dimensional
space, he finds the same result. Therefore, his finite accuracy is consistent
with particle trajectories which are separated from one another by the action
of what appear to be loops (enclosed regions) when he projects his data onto a
hyperplane. These loops are a measure of the possible differences between the
particle histories. The observer also cannot determine whether these
loops are linked or not, and again concludes that both situations can occur.
Therefore, even though the particle trajectories through space-time may very
well be topologically trivial, the
observer has no way of confirming this possibility.

Imagine the observer in a satellite orbiting a black hole. To determine the
properties of the black hole and to see if his conclusions about particle
trajectories differ from the zero gravity case, he first establishes the
existence of space-time curvature.
The observer then measures the motions of probes in closer and closer orbits
and establishes the existence of an event horizon. When he performs
measurements within $\ell$ of the Schwarzschild radius, he again concludes
that there is an uncertainty as to the probe's actual trajectory. Even though
the observer is disturbed by the fact that he cannot rule out that some probes
moved into and out off the event horizon, he assumes that higher accuracy
measurements will. The observer can conclude that there is a closed surface and
only one generic loop associated with single particle trajectories
due to the focusing effect of space-time curvature close to the horizon.

Now {\it postulate} that the finite accuracy of the observer's measuring rod
is a property of Nature itself when $\ell =\ell_{\rm Planck}$, i.e.\ Nature
reaches these same conclusions {\it everywhere}.
The measurements then reflect an {\it intrinsic} property rather than an
observational external one, and the possible existence of topological
obstructions in the form of loops
becomes a {\it requirement} to explain the possibility of
different particle trajectories. Also {\it use} a mathematical framework
in which Planck scale space-time possesses the property of continuity, and is
locally flat when gravitational effects can be ignored. One then arrives at a
description where quantum mechanics and general relativity require
the presence of the flat three-torus, which introduces a length scale
$\ell_{\rm Planck}$, and the curved handle, which gives in addition a mass
$m_{\rm Planck}$ (black holes can come in all masses {\it and} sizes).
The non-trivial loop homotopy of these manifolds now facilitates
the possible distinct particle trajectories. It
follows that the observed loops are {\it dynamical} objects which reflect a
fundamental uncertainty in Nature. Note also that this suggests that
the maximum combined spatial and temporal dimension of the universe
is four since the possible linkage of loops is a topological invariant in
three dimensions only.

One can now provide a physical basis for Mach's principle.
The question is how a particle knows which way to move under the influence
of inertial forces, if one rejects the notion of absolute space-time.
The thought experiment suggests that it is the combined motions of all
particles up to the present. Since these motions are an expression of the
curved lattice of three-tori with handles, the geometry of space-time as well
as its topology determines how particles move\footnote{Note that the different
trajectories contributing to the Feynman path integral now reflect the
topological freedom in space-time itself.}.
Because the thought experiment applies to any time-like slice and for {\it all}
of the projected loops therein, it is the global topology (just like it is the
global geometry for Einstein gravity) which enters Mach's principle.

The prime quantum of $T^3$ under $O$ then reflects the fact that in a lattice,
any point (where particles are moving) is connected to six others (where other
particles are moving), through three-ball surgery.
In the case of $S^1\times S^2$, the topology of the two-sphere only admits
two such paths. All these manifestations of the prime manifolds are
{\it physical} because it is the {\it combined} motion of particles which
determines their inertial frames. Because the three-tori only possess
a scale, their prime quantum represents an effective dimension of seven.
For the massive handles one concludes that each induces (irrespective of
position and size) the spontaneous creation and evaporation of two mini black
holes every Planck time, and that they constitute the cosmological constant on
the left hand side in the Einstein equation.

\subsection{Mathematical Formulation}

After these arguments, one can arrive at the underlying mathematics through
the following line of thought. The equivalence principle in general relativity
leads to the notion of gravitational accelerations as
an expression of space-time curvature. In quantum theory the
concept of a force is equivalent to an interaction between fields
and the notion of a curvature two-form generated by some gauge potential.
Mathematically, a curvature involves the second derivatives of
some metric field. The occurrence of a second derivative of any field
requires that field to be defined in three points. These three points need not
be infinitesimally close together because the coarseness of a derivative only
depends on the measurement scale one is interested in. This leads to a natural
partition of space into triplets. That is, the concept of a force field in a
mathematical model requires the possibility of any triplet of points to be
mapped into a curvature.

This yields a formal object $s$ denoted as
$s=\{\cdot\cdot\cdot\}\equiv [123]$, where the three dots indicate
the notion of a second derivative defined by three arbitrary points.
The topological structure reflected by $s$ follows from the realization that
the paths (partial trajectories) connecting [1], [2] and [3], are distinct.
This then implies
$$[12]\ne [23]\ne [13],\eqno(1)$$
and
$$[ab]=[ba],\eqno(2)$$
where the last relation reflects that there is no preferred orientation.
Clearly, these relations imply the homotopic structure of the three-torus with
the loops $[12]$, $[23]$ and $[13]$, {\it if} one demands that the three pairs
of points cannot be taken infinitesimally close together. The handle and
three-sphere follow analogously. When these loops are viewed as dynamical
objects, their creation and annihilation operators clearly do not commute,
$[T,T^\dagger ]=1$.

\section{The Fundamental Manifold and the Equation of Motion}

\subsection{QGUT Phenomenological Preliminaries}

One should first assess which types of interactions the combination of $T^3$
and $S^1\times S^2$ allows from a {\it purely topological} point of view.
The three-torus has 3 loops and it is easy to see that individual loops yield
spin 1 particles and pairs of loops lead to spin 1/2, through the introduction
of an internal angle $\gamma$ which runs between $0$ and $2\pi$ for each
loop[1,4]. Clearly, spin 1/2 as a fundamental two-valuedness occurs naturally,
and $T^3$ supports both submanifolds without preference. The handle introduces
an additional degree of freedom which can increase or decrease the homotopic
complexity of the structure to which it is attached. Since the handle reflects
the gravitational interaction, one can conclude immediately from the thought
experiment that locally any $T^3$ loop must be connected to the mouth of a
handle {\it if} gravitational and non-gravitational forces are unified {\it
and} the universe is in its ground state, i.e.\ black holes maximize entropy
and each path through space-time can enter an event horizon. The aim is
now to find a three-dimensional manifold consisting of three-tori and handles,
which reproduces the standard model through this homotopic implementation of
unification.

Since this constructed manifold will be the building block for a lattice,
one should also consider symmetries generated by three-ball surgery.
As one patches the individual three-tori
together, there is an arbitrary $O(2)$ rotation, or twist, one can
perform without changing the topological properties of the manifold. This
$U(1)$ gauge freedom in the lattice junctions allows for a supersymmetry
because one can perform arbitrary rotations of $\gamma$ through the junctions
in $L(T^3)$, thus facilitating a change in fermionic or bosonic spin structure.
This symmetry will be referred to as a lattice supersymmetry.

In paperI it is shown that the $T^3$ lattice junctions support $SO(n)$ and
$SU(n)$ symmetry groups because $L(T^3)$ naturally yields quadratic and quartic
interaction terms and a self-interaction potential
$V=\mu^2\Phi^2+\lambda\Phi^4$, for constants $\mu$, $\lambda$, and a
scalar\footnote{The vacuum expectation value of $\Phi$ must be Lorentz
invariant because of the three-torus.} multiplet $\Phi$. The maximum degree of
four in the interaction potential follows from the fact that there are four
homotopically inequivalent paths on $L(T^3)$. The potential $V$ is even since
the field $\Phi$ is defined on $S^1$ loops.
With the prime quantum of the fundamental manifold $T^3$ discussed above, it
follows that the dimension of the configuration space defined by $L(T^3)$
is $D=7+3+1=11$. The homotopically trivial
manifold $S^3$, which is needed in the construction
of $L(T^3)$, and the macroscopic handle provide the large scale
(compactified) topology of the universe. The
junctions which connect the three-tori correspond to the propagators in a
field theory and are homotopically equivalent to line elements. It follows
that the dimension of the junction groups which describe the interactions
between the propagators on $L(T^3)$ is then $11-1=10$ at any time, i.e.\
$SU(5)$ and $SO(10)$, which are viable candidates for a GUT. The very presence
of the three-torus also hints strongly at an underlying link with M-theory and
the T-dualities among superstrings.

\subsection{QGUT Construction}

\subsubsection{The Fundamental Topological Manifold $Q$}

A manifold $Q=aT^3\oplus bS^1\times S^2$ which is built from three-tori and
handles should have an odd number of
constituents because only odd sums of nuclear primes bound Lorentz manifolds
(paperI). Because the unification of forces is
implemented homotopically, the number of loops in the $a$ three-tori should be
equal to the number of mouths of the $b$ handles, i.e.~$3a=2b$.
Finally, any solution which has $a_1=na$ and $b_1=nb$ can be considered a
multiple of the smallest solution $aT^3\oplus bS^1\times S^2$. Since one wants
to construct a lattice $L(Q)$, the minimal solution is the desired one.
Therefore, the coefficients $a=2$ and $b=3$ result. This requires the direct
sum of three handle manifolds and two three-tori. In $Q$, each handle is
connected to two loops and the ground state manifold can be written as
$$Q=2T^3\oplus 3S^1\times S^2,\eqno(3)$$
which assures nuclearity and therefore Lorentz invariance. When the density of
charges is large enough to support $Q$, the universe is said to be
$Q$-symmetric.

\subsubsection{The Equation of Motion for Constant Charge}

In order to develop a field theory which involves matter interactions at the
Planck energy, the topological object $Q$ is considered to be fundamental.
The equation of motion for the wave function should then follow from some
continuum limit of the loop algebra acting on the four-manifold $Q\times R$.
The natural limit of the loop creation and annihilation operators
is that of two differential operators $\partial$ and $\partial^\dagger$
with space-time dimension four. $\partial$ and $\partial^\dagger$ must commute
in the continuum limit because $[T,T^\dagger]=1$ and the 1 on the right hand
side reflects the discrete nature of the loop algebra. These differential
operators should also be conjugate in order to yield a scalar operator of the
form $\Box =\partial\partial^\dagger =\partial^\dagger\partial =
\partial_\mu\partial^\mu$, $\mu =1..4$.
These considerations require the identification
$(T,T^{\dagger}\rightarrow\partial_\mu ,\partial^\mu)$.

The interactions present in the equation of motion {\it must} follow directly
from the topological structure of $Q$ and $L(T^3)$ {\it if} one accepts the
thought experiment. One demands on the left hand side a single index equation
because there are four homotopically
inequivalent paths on $L(T^3)$; cubic interactions since every loop is attached
to a handle and a junction; scalar quadratic interactions due to individual
loops; and linear second derivatives
(curvature). The right hand side is zero because $Q$ and $L(T^3)$ are compact.
This yields for the complex four vector $q_\lambda$
$$q^\mu [\partial^\nu q_\mu ,\partial_\nu q_\lambda ]=0,\eqno(5a)$$
where only the commutator form satisfies the right hand side for solutions of
the Klein-Gordon equation. One finds
$$q^\mu [q_\lambda\Box q_\mu -q_\mu\Box q_\lambda ]=0.\eqno(5b)$$

The square of the absolute value of the wave function,
$q^\mu q_\mu\equiv \delta^{\mu\nu}q^*_\nu q_\mu$, assures a positive
definite inner product and a well defined probability.
The energy of $Q$ is given by the dimensional number
$m_{\rm Planck}$, generated by the handle triplet\footnote{The third
dimensional number in the theory is the speed of light.}.
The individual components of $q_\lambda$ yield probabilities for each of
the four homotopically distint paths through $L(T^3)$. In PaperI it is shown
that the $SO(3,1;R)$ gauge group follows naturally from a nuclear manifold.
The theory is therefore manifestly Lorentz invariant with
$\partial^\mu =\eta^{\mu\nu}\partial_\nu$ and $\Box$ the d'Alambertian.

\subsubsection{The Equation of Motion in the Presence of Charge Fluctuations}

For large charge densities, $L(T^3)$ is completely
interconnected by handles. In this limit $L(T^3)\rightarrow L(Q)$ as the Planck
scale manifold. Unlike the three-tori, the mini
black holes couple directly to the matter degrees of freedom through the
process of Hawking radiation. Therefore, even though
the charge density is of the order of unity, handles are continuously being
created by the global distribution of handles ($\Lambda$) and destroyed through
evaporation (paperI).

These quantum perturbations in the local number of handles lead to the
generation of an additional field. This field is envisaged to reflect phase
changes (the topology of $Q$ is not altered) in the wave amplitudes
$q_\lambda$, flowing through $L(T^3)$. The fundamental object to solve for on
$L(Q)$ is therefore $\Omega_\lambda\equiv {\rm e}^{2\pi i\phi}q_\lambda$, with
$\phi$ a function of time and position. This phase transformation leads to the
full QGUT equation of motion
$$4\pi i\partial_\nu\phi [(\partial^\nu q_\lambda )q^\mu q_\mu -(\partial^\nu
q_\mu )q^\mu q_\lambda ]=q_\lambda q^\mu\Box q_\mu -q^\mu q_\mu \Box
q_\lambda ,\eqno(6)$$
with an additional scalar constraint
$$q^\mu q_\mu ={\rm cst},\eqno(7)$$
which becomes void when $Q$-symmetry is broken. The scalar constraint signifies
the fact that, due to the continuous creation and destruction of handle
manifolds, it is possible to travel from one point along a homotopic path to
any other point along a different homotopic path in $Q$. Therefore, from the
perspective of the wave amplitudes, any point in $Q$ becomes indistinguishable
from any other, while the topology of the thought experiment persists.
Equation (7) thus expresses that the {\it total} probability to be somewhere
in $Q$ is the same for all points in $Q$.

Because the evolution of $\phi$ is driven by the handles,
it follows that on the neutral $T^3$ lattice the field $\phi$ obeys the
limiting condition $\partial_\nu\phi =0$, and is effectively frozen in at
a value which does not have to be zero a priori.
In the zero charge limit, the numerical value of the field $\phi$ must
correspond to a constant, not necessarily zero, Lorentz invariant vacuum
expectation value, i.e.\ $<0|\Phi |0>$ under the junction potential $V$ on
$L(T^3)$. A non-zero vacuum expectation value of $\Phi$ would require
$\mu^2<0$ in $V$ and can lead to
spontaneous symmetry breaking, as first suggested by Nambu and co-workers.
The additional scalar $\phi$ therefore induces a Higgs field
$\Phi$, although the number of Higgses is not constrained. In Section 4, the
topology of $Q$ will be used to fix the end value of $\phi$ as well as the
vacuum energy associated with $V$.

\subsubsection{The Thermodymanics of $Q$}

The number of degrees of freedom of $Q$ under the action of the loop algebra
operator $O$ is defined as
$$N_QQ=(TT^\dagger +T^\dagger T)(2T^3\oplus 3S^1\times S^2)=23Q.\eqno(8)$$
In the loop homotopic approach adopted here, these degrees of freedom are
all distinct and they reflect the different ways in which the loop creation
and annihilation operators can act on $Q$, i.e.\ induce the different possible
topological obstructions from the thought experiment. As discussed above, the
weight of 23 which $O$ assigns to $Q$, represents the actual number of Planck
scale realizations of the manifold at any given time. Because $Q$ is the
topological expression of a mass, a scale, {\it and} a gauge group, the
degrees of freedom of $Q$ are identified with different particle states.

For the neutral submanifold $P=T^3\oplus T^3$,
one has $N_P=14$. The latent heat associated with the evaporation of the
handle triplet $\Theta =3S^1\times S^2$ is therefore
$$H=(N_Q-N_P)m_{\rm Planck}/N_Q=9m_{\rm Planck}/23.\eqno(9)$$
This number is uniquely determined by the homotopic structure of space-time
and the Planck mass. Since the two three-tori in the structure $Q$ are
identical objects, the specific heat per three-torus is given by
$$h=H/2=9m_{\rm Planck}/46.\eqno(10)$$

\subsubsection{The QGUT Phase and Large Scale Structure}

During the $Q$-symmetry phase the 23 degrees of freedom of $Q$ are accessible
to the wave amplitudes $\Omega_\lambda$ traveling through $L(Q)$. As these
currents self-interact, they do so cubically as in equation (6).
The amplitudes carry the mass-energy of the universe and their interactions
determine the perturbations in that mass-energy, which is distributed over the
23 realizations of $Q$. The equation of motion (6) is invariant under a
global scale transformation $q_\lambda\rightarrow Aq_\lambda$, and the
topology of $Q$ then fixes $A=N_Q^{-1}$.

The magnitude scale of the perturbations in the mass-energy is thus given by
$$\delta\rho /\rho(Q) =N_Q^{-3}=8.2\times 10^{-5},\eqno(11a)$$
where the degrees of freedom are considered equally accessible
thermodynamically. For adiabatic perturbations, the fluctuations in the Cosmic
Microwave Background (CMB) temperature are $1/3$ of $\delta\rho /\rho(Q)$ and
one finds a 1 $\sigma$ Gaussian (by assumption) CMB amplitude
$$\delta T/T={{1}\over{3}}\delta\rho /\rho(E) =2.7\times 10^{-5}.\eqno(12)$$
This value is consistent with recent COBE and ground based measurements,
although it does not include any physics after QGUT symmetry-breaking.
Analogously, one finds that the characteristic dispersion of $T^3$ is $7^{-3}$.

\section{The Symmetries of $Q$ and $L(T^3)$ in the Standard Model}

With the fundamental manifold and the equation of motion for the wave
function which lives on it in place, one now needs to isolate the symmetry
properties of $Q$ and $L(T^3)$ in order to establish a link with the standard
model, and to provide an {\it interpretation} for the equation of motion in
terms of matter properties on $L(T^3)$.

\subsection{General Properties of $Q$}

\subsubsection{Discrete Groups Generated by $P$ and $\Theta$}

The handle manifolds which are created as quantum fluctuations form
triplets on $L(Q)$. Therefore, the effective action $s^3$, of the
triplet as a whole, obeys
$$s^3=1.\eqno(13)$$
That is, a round trip along the manifold $\Theta$ necessarily picks up three
phases, which should add up to $2\pi$ since the loop algebra satisfies
$[T,T^\dagger ]=1$. Because all 3
handles are identical, this implies a $Z_3$ invariance for the individual
quantum fields in the theory defined on $Q$, with angles
$\theta_i=\{0,\pm 2\pi /3\}$. From the same arguments it follows that the
submanifold $P=T^3\oplus T^3$ in $Q$ generates a $Z_2\times Z_3$ symmetry
because one cannot distinguish either three-torus in $P$.

\subsubsection{$T^3$ Junctions and $U(1)$ Symmetries}

The important distinction between $T^3$ and $L(T^3)$, or $L(Q)$ for that
matter, is the presence of junctions which connect the individual three-tori
through three-ball surgery and create a lattice. In paperI it was suggested
that the presence of a lattice facilitates a geometric description of
gravitational effects because the three-dimensional junctions can bend
according to some curvature tensor. Indeed, on scales much larger than
$\ell_{\rm Planck}$ the neutral lattice $L(T^3)$ appears as a smooth manifold.
On the Planck scale on the other hand, the existence of junctions between the
three-tori generates an additional $U(1)$ symmetry as discussed above.
Obviously, there is only one twist per $Q$ manifold, but each individual
three-torus in the lattice formally has six of them. Such additional $U(1)$
factors have been proposed as a possible resolution of the doublet-triplet
splitting problem[5], and will be studied below.

\subsubsection{Particle Sectors on $Q$}

The homotopic properties of $Q$, the $SU(5)$ or $SO(10)$ gauge
group on the junction,
and the $Z_3$ and $Z_2\times Z_3$ cyclic symmetries of $\Theta$ and $P$ should
lead to specific particle sectors. In this, the photon and graviton are not
viewed as being generated through the homotopic structure of $Q$, but result
from the junction degrees of freedom, i.e.~the $U(1)$ twist and $GL(4)$
curvature of the lattice. Furthermore, there is a
Higgs field $\Phi$ which lives on the lattice and can cause symmetry
breaking, leading to Higgs bosons.

The number of degrees of freedom $N_Q$ is the eigenvalue of the operator
$O=TT^\dagger +T^\dagger T\equiv A+B$ acting on $Q$.
There is then a natural division of the 23 degrees of freedom under
$AQ=14Q$ and $BQ=9Q$. Furthermore, the decomposition $OQ=O(P\oplus\Theta )$
has the same distribution of degrees of freedom under $A$ and $B$ and leads to
the further divisions
$$AQ=OP=(8+6)P,\eqno(14)$$
with eight plus six particles and
$$BQ=O\Theta =(3+6)\Theta ,\eqno(15)$$
with three plus six particles. The number of loops plus the number of pairs of
loops tells us that there are (6+6) elementary particles. The number of field
particles then follows naturally from the decomposition.
The junction potential on $L(T^3)$ or $L(Q)$ supports the symmetry group
$SU(5)$ or $SO(10)$. The $P$ and $\Theta$
sectors decompose $Q$ and are therefore associated with subgroups.
These subgroups can only contain $SU(n<5)$ and $U(1)$ because of the junction
potential $V$ and twist. For $SU(5)\sim SU(3)\times [SU(2)\times U(1)]$ these
constraints are satisfied, because $SU(3)$ contains 8 ($P$) field particles
and $SU(2)$ only 3 ($\Theta$). Below it will be shown that the field particles
must be bosonic. Therefore, $Q$ defines a ground state which
corresponds to the standard model with only one $U(1)$ factor.

\subsubsection{Supersymmetry Breaking and the Pauli Exclusion Principle}

A priori, both fermionic and bosonic sectors exist for the equivalence
classes identified above. That is, because the form of $Q$ is motivated by
Lorentz invariance and unification, the identified equivalence classes can be
both fermionic and bosonic in nature.
When $Q$-symmetry is broken, the lattice supersymmetric structure $L(T^3)$
becomes the fundamental (Lorentz invariant) Planck scale object. Subsequently,
interactions are mediated by field particles which travel along the 6 junctions
surrounding any $T^3$. That is, it is the discrete three-torus with its
seven degrees of freedom under the operator $O$ which supports a field and
its quanta. Any field dynamics on $L(T^3)$, {\it after} lattice
supersymmetry has been broken, therefore requires the interaction of two
(identical) field particles on a three-torus. If these field particles are
manifestly fermionic, this violates the Pauli exclusion principle. Thus, only
bosonic field particles can carry the strong and electro-weak force, and
satisfy the Pauli exclusion principle on $L(T^3)$, after supersymmetry has
been broken.

The origin of the Pauli exclusion principle actually {\it follows}
from the homotopic structure of $T^3$ and the fact that a spin $1/2$
particle requires two loops on a three-torus for its support. For two identical
fermions one finds the general relation (see \S 2.2 above)
$$[ac][cb]=[ca][ab],\eqno(16a)$$
which yields
$$[cb]=[ab].\eqno(16b)$$
This indicates that because one $S^1$ loop is a part of both fermions, the
other two are collapsed to one. The consequence is that the three-torus becomes
indistinguishable from the prime manifold $S^1\times R_1$, with $R_1$ a
Riemann surface of genus one. This manifold is not nuclear and breaks Lorentz
invariance. Therefore, one finds a {\it topological} restriction which
precludes interactions mediated by fermionic field particles. Any
super-partners of fundamental particles are therefore rendered inert, except
for gravitational interactions.

\subsection{Interpretation of the Equation of Motion}

Equations (5), (6) and (7)
describe the quantum-mechanical interactions of matter in full,
i.e.~including quantum gravity, without any need to know the specific
properties of the particles in the ultimate field theory.
The boundary conditions for the solutions to these equations
follow from the cyclic properties of $L(T^3)$.
The topology of $T^3$ requires the solutions $O_\lambda (x,y,z,t)$ on the
lattice to be periodic on a scale $L_i=n_i\ell_{\rm Planck}$ for positive
integers $n_i$, $i=1..3$, and at every time $t$,
$$O_\lambda (x,y,z,t)=O_\lambda (x+L_1,y+L_2,z+L_3,t).\eqno(17)$$
The $Z_3$ group of the handle triplet then yields a solution which limits to
$|\phi |=1/3$ or 0 when $Q$-symmetry is broken. This phase relation is
consistent with the fact that the equation of motion (6) is invariant under the
global transformation $\phi\rightarrow\phi +\alpha$. This freedom is thus
fixed by the underlying topology of $Q$. The initial conditions at $t=0$ for
the solutions of (6) can then be taken as $q_\lambda (0)={\rm cst}$,
derivatives $\partial_t q_\lambda (0)=1$ in Planck units, and $\phi (0)=0$.

Note here that $\phi$ is driven by the $\Theta$ manifold, and can lead to
inflation through the vacuum energy associated with the interaction potential
$V$ {\it if} it limits to a non-zero end value. Most importantly, the possible
final values of $\phi$ after $Q$-symmetry breaking are determined by the $Z_3$
symmetry of $\Theta$ and lead to different solutions of the equation of motion,
thereby naturally {\it including} the influence of this vacuum energy. Below it
is shown that this vacuum energy (the GUT scale) is uniquely determined by the
topology of $Q$, so that the equation of motion knows about it. This provides
a crucial link between the mass-energy wave function and the observational
characteristics of the low energy field theory.

To follow the evolution of the wave function after the handles have
evaporated\footnote{If merging contributes significantly, then a lot of
primordial black holes may be formed.},
one should solve equation (5) with the end solution of (6) as initial
conditions. During this phase, the characteristic amplitude of the
fluctuations remains as computed above because
all the original 23 degrees of freedom of $Q$ (later to become particles) are
above the GUT unification scale (see its computation below). Once the GUT is
broken at some energy, the Einstein equation describes the later time
evolution of the mass-energy distribution (the expectation value of
$q^\mu q_\mu$), as the universe expands. The fact that general relativity does
not constrain the topology of space-time thus appears to follow from the fact
that it is only valid if one can ignore the topology of space-time.
Nevertheless, Einstein gravity is still a part of the QGUT through the
presence of the handles.

A question which can be addressed through solutions to (5) and (6) is the
nature of the statistics of {\it mass-energy} fluctuations during the
radiation-dominated era, after the GUT is broken. Furthermore, the solutions of
(5) on $L(T^3)$, with the dispersion $7^{-3}$, then provide the possible
probability distributions for the {\it rest masses} of particles.
Finally, the periodicity condition (17) renders $O_\lambda$ identical on each
$T^3$ prior to GUT breaking. Therefore, if inflation increases the size of a
Planck scale region to no more than the local horizon scale, then periodicities
in the matter distribution could be present at the current epoch.

\subsection{Predictions for the Standard Model}

\subsubsection{Unification Energies on $L(Q)$ and $L(T^3)$}

A fundamental problem which requires a resolution in GUT is the specific
form of the fermion mass hierarchy. The popular approach is to use the
anomalous $U(1)$ gauge symmetry as a horizontal symmetry[5]. The
motivation is that the $U(1)_{\rm A}$ symmetry breaking scale is given by
the square root of the Fayet-Iliopoulos term $\xi$[5]. The ratio of $\xi$ to
the Planck mass is of the order of the fermion mass ratios in neighboring
families. The precise value of $M_{\rm A}=\surd \xi /q$, with $q$ the
anomalous U(1) charge, depends on the value of
trQ=trQ$_{\rm obs}$+trQ$_{\rm hid}$, with contributions from the observable and
hidden matter singlet[5]. A second mass scale corresponds to $M_{\rm GUT}$,
which marks the energy at which the $SU(5)$ or $SO(10)$ theory is broken down
to the $SU(3)\times [SU(2)\times U(1)]$ symmetry of the standard model.
Obviously, both energy scales should have a common origin.

When the handles evaporate, the value of $|\phi |$ can be $1/3$ which breaks
the $SU(5)$ or $SO(10)$ symmetry through $V$. Since $Q$ is the ground state
for a grand unified theory, the GUT energy scale should correspond to
$1/3$ ($Z_3$ introduces three branches) times the energy of a degree of freedom
per three-torus ($P=2T^3$) on $Q$. Because the 23 degrees of
freedom on $Q$ are equally favorable one expects equipartition of energy.
Therefore, one finds $M_{\rm GUT}=1/3m_{\rm Planck}/2N_Q= 8.8\times 10^{16}$
GeV, with $m_{\rm Planck}=1/\surd G=1.22\times 10^{19}$ GeV in units with
$\hbar =c=1$. Above it was shown that the latent heat per $T^3$ associated with
breaking of $Q$-symmetry, which yields $L(T^3)$, equals
$h=9/46m_{\rm Planck}$. Therefore, one finds
$M_{\rm A}=1/3h= 8.0\times 10^{17}$ GeV, for the energy above which lattice
supersymmetry operates\footnote{Note that on $L(T^3)$ the GUT and
$U(1)_{\rm A}$ symmetries are broken at an energy {\it per degree of freedom}
which is a factor of 7 smaller than $M_{\rm GUT}$ and $M_{\rm A}$.}.

Finally, the energy per degree of freedom on $Q$ is $m_{\rm Planck}/23$ whereas
it is $m_{\rm Planck}/7$ on the three-torus because $OT^3=7T^3$. Therefore,
when the handles evaporate the energy per degree of freedom on $L(T^3)$
increases by a factor $23/7$. It follows that $Q$ is the ground state in an
entropy {\it as well as} energy sense. From the above one then finds that the
transition $L(Q)\rightarrow L(T^3)$ leaves the universe in an {\it excited
state} associated with the supersymmetric extension of $SU(5)$ or $SO(10)$
through $U(1)_{\rm A}$, whose decay can leave behind new particle states.

\subsubsection{$Q$ Symmetry Groups and Doublet-Triplet Splitting}

Another fundamental problem in supersymmetric GUT is the doublet-triplet
splitting problem which results from the unavoidable mixing of Higgs doublets
$H,\bar{H}$ with their colored triplet partners $T,\bar{T}$.
This also leads to an unacceptably rapid proton decay. In [6] it was suggested
that there is no need for the heavy triplet if its Yukawa coupling constant is
strongly suppressed with respect to the one of the doublet. This mechanism
requires an SO(10) invariant operator with tensor indices $i,k$
$${{Y_{\alpha ,\beta}}\over{M_{\rm GUT}}}10_i45_{ik}16^\alpha
\gamma_k16^\beta ,\eqno(18)$$
in which $16^\alpha$ ($\alpha =1..3$) are three families of matter fermions,
$10_i$ ($i=1..10$) is the multiplet with $H,\bar{H}$ ($i=7..10$) and
$T,\bar{T}$ ($i=1..6$). The 45 is the GUT Higgs in the adjoint presentation of
SO(10), $Y_{\alpha ,\beta}$ is the coupling constant matrix, and the $\gamma_i$
denote the matrices of the SO(10) Clifford algebra. To realize this effective
operator, the 10-plet must transform under the symmetry group $Z_2\times Z_3$
so that it does not couple to the GUT Higgses and is
allowed to interact with $16^\alpha$ only in combination with the 45-plet[6].

A possible resolution of the doublet-triplet splitting problem in the standard
model thus naturally involves the cyclic group $Z_2\times Z_3$ associated with
$P$. If one now introduces a light gauge singlet superfield $N$, then the
triple interactions on $L(T^3)$ {\it determine} the associated interaction
potential to be of the form\footnote{Recall that (6) has a {\it fourth} order
interaction term on the left and a {\it third} order interaction term on the
right hand side.}
$$W_\mu =\lambda_1 N10^2+\lambda_2N^3.\eqno(19)$$
This potential is automatically invariant under $Z_2\times Z_3$ because the
singlet $N$ has an $Z_2$ invariance on $P$. Both $N$ and 10 do not transform
under $Z_3$ due to $\Theta$, and therefore they decouple from the heavy GUT
Higgs fields. This provides a natural resolution of the $\mu$ problem in terms
of the homotopy of space-time through the symmetry groups anticipated in [6],
{\it if} one accepts the existence of an additional gauge singlet. Because the
triplets have no coupling at all, it follows that the proton is essentially
stable even if the decoupled triplet is as light as its doublet partner. That
is, its decay rate is suppressed by a factor which is no larger than
$(M_W/M_{\rm GUT})^2$, with $M_W$ the mass of the weak scale. Specific models
based on SO(10) are constructed in [6].

The existence of the supersinglet should be associated with a
dynamical symmetry, just like $\phi$ reflects $Q$-symmetry. The only other
symmetry related to the $L(Q)\rightarrow L(T^3)$ transition is the anomalous
$U(1)$ symmetry, i.e.\ lattice supersymmetry, which is suppressed in $L(Q)$ but
emerges on $L(T^3)$. The additional singlet is therefore associated with
$U(1)_{\rm A}$ and is supersymmetric. Its characteristic energy
breaking scale is the topological number $M_{\rm A}$ computed in \S 4.3.1.

\subsubsection{The Absolute Scale of the Mass Ground State}

It will now be shown that, just like the energy scales of the
fields generated by $Q$ are uniquely determined by the homotopic theory, so is
the low mass end of the lepton mass hierarchy in the electro-weak
$\Theta$ sector. Conversely, no such limits can be placed on the QCD sector,
except through direct relations between different generations in terms of the
horizontal $U(1)$ symmetry on $L(T^3)$.

Before $Q$-symmetry is broken, the 23 degrees of freedom of $Q$ are in
equilibrium and there are $X_Q=23!$ different configurations on $Q$. The mass
of the particle ground state is thus $m_{\rm Planck}/X_Q$. The particle should
be charged because $Q$ contains a $U(1)$ sector. It follows that
$$m_{\rm e}^0=m_{\rm Planck}/X_Q= 0.47\quad {\rm MeV},\eqno(20)$$
determines the electron mass.

For the neutral submanifold $P$ one has $X_P=14!$, which fixes the neutral
ground state on $L(T^3)$. Because the neutrinos have no charge, they cannot be
distinguished on $P$, unlike the charged leptons which couple to the $U(1)$
sector on the junction of $Q$. The total number of configurations is now
$X_QX_P$. The mass of the neutral (electron neutrino) ground state thus follows
from
$$m_{\nu_{\rm e}}^0=m_{\rm Planck}/(X_QX_P)= 5.4\times 10^{-6}\quad
{\rm eV}.\eqno(21)$$
The photon and graviton are massless because they do not depend on the homotopy
of $Q$. The gluons remain
massless because they are associated with $P$. The masses of the vector bosons
are determined by the breaking of the $SU(2)\times U(1)$ gauge symmetry through
the $|\phi |=1/3$ solutions, which requires direct intervention by $\Phi$.

\subsubsection{Corrections to the Mass Ground State}

If $s_Q=23^{-3}$ describes the dispersion of the probability distribution
on $Q$ (computed above for the CMB), then one can ask with what accuracy
${\cal A}$ the properties of $P\subset Q$ can be determined, given that it has
a dispersion $s_P=14^{-3}$. The uncertainty relation then yields
$s_Q={\cal A}s_P$. This question is relevant to the particle mass in the
charged and neutral state since the handles occupy only a part of the total
number of degrees of freedom on $Q$. The relative uncertainty in the ground
state masses is therefore ${\cal A}=(14/23)^3=0.23$. This uncertainty reflects
an upward shift in the mass because the finite accuracy ${\cal A}$ implies an
intrinsic lack of information. From the $Z_3$ symmetry of $\Theta$, one finds
that the magnitude of the shift is ${\cal A}/3$ which yields
$$m_{\rm e}=(1+{\cal A}/3)m^0_{\rm e}=0.51 \quad {\rm MeV}.\eqno(22)$$
This is in remarkable agreement (error $<0.7$\%) with the measured value of
$0.511$ MeV, and lends support to the notion that the 23 degrees of freedom of
$Q$ are truly fundamental. Analogously, one finds that
$$m_{\nu_{\rm e}}=(1+{\cal A}/3)m_{\nu_{\rm e}}^0=5.8\times 10^{-6}\quad
{\rm eV}.\eqno(23)$$
The fact that the electron neutrino has a mass is a {\it unique} prediction of
the model.

\subsubsection{CP Violation}

There is one more degree of freedom on $L(T^3)$. The lattice junction of a
pair of three-tori can vibrate. This vibration is such that no curvature is
induced, like the motion of a piston, motivated by the thought experiment. The
energy $F$, corrected for ${\cal A}$/3, of the excitation is
$$F=(1+{\cal A}/3){{m_{\rm Planck}-H}\over{X_QX_P}}=3.54\times 10^{-6}
\quad {\rm eV.}\eqno(24)$$
The denominator reflects the total number of neutral configurations on $Q$.
Until the mini black holes evaporate and the latent heat $H$ is released, an
energy $m_{\rm Planck}-H$ is confined to the internal degrees of freedom of the
three-tori. Of course, the presence of this state reflects the conceptual
difference between the $\Theta$ and $P$ sector on $Q$. This vibration, if
excited, is a special one in that it violates T invariance. That is, the
Planck time vibrations result from a homeomorphism, not a diffeomorphism, of
space-time.

To excite this internal degree of freedom one requires a system of two neutral
particles interacting through a common decay route. The energy difference
between the rest masses of their superposition states should be equal to a
positive integer multiple $m$ of $F$ (resonance) in
order to excite the vibration. It is well known that the difference in weak
self-energy determined by the superposition states
$$K_S\leftrightarrow 2\pi\leftrightarrow K_S\eqno(25a)$$
and
$$K_L\leftrightarrow 3\pi\leftrightarrow K_L,\eqno(25b)$$
for the decay of the $K^0$ and $\bar{K}^0$ mesons, is extremely small and
equal to $f=3.52\times 10^{-6}$ eV. Indeed, $f\approx F$ to less than one
percent. It is this remarkable coincidence which can allow the CP violating K
meson decay processes to occur through $K_L\rightarrow K_S\rightarrow 2\pi$.
This conclusion demands the validity of CPT invariance. The possibility of a
CP violating process results from a vibration in space-time itself, which
therefore violates T invariance. Now if CPT is preserved, then T violation in
fact {\it requires} the CP violating decay of the K meson. Although it is not
clear whether CPT should be preserved under all circumstances (it is for $K$
decay), the point of view is taken here that it always is.

The level of violation (experimentally $0.227$\% for two-pion decay) is not
predicted by this argument, but is related to the excess in $F$ over $f$. If
$F$ is increased by $0.63$\% to yield perfect agreement with the measured
electron rest mass, i.e.\ our surroundings are characteristic of a
$\sim 2\sigma$ perturbation in the wave function, then the relative energy
difference between $F$ and $f$ is $E=1.29$\%. For the intrinsic dispersion
$d=7^{-3}$ of the $T^3$ lattice vibration and an exponential decay rate
$\Sigma_n[7^{-m}{\rm exp}^{-E/d}]^n$ per total number of realizations of the
three-torus and summed over all resonances, one finds $R_{\rm CP}=0.17$\% for
$m=1$. This is in reasonable agreement with experiment, given the use of a
simple exponent for the barrier penetration.

\section{Conclusions, Discussion, and Future Prospects}

A thought experiment has been proposed which leads to the notion of
three-tori and handles as true fundamental objects on the Planck scale,
embodying the interplay between general relativity and quantum mechanics.
Together these prime manifolds form a fundamental topological manifold which
is Lorentz invariant and provides a natural mechanism for symmetry breaking.
The general equation of motion has been derived for a possible QGUT on this
manifold $Q=2T^3\oplus 3S^1\times S^2$, which naturally leads
to a Higgs field (driving inflation) and the amplitude of the primordial
density fluctuations. The manifold
$Q$ contains the necessary symmetry groups to reproduce the standard model as a
ground state with a stable proton, and possesses intrinsic energy scales which
determine the masses of the lightest leptons. An attractive feature of the
constructed theory is the natural (3+1) dimensions, without the need for
compactification. The presented model can be falsified immediately by
measurement of the electron neutrino mass, CP violation, and by comparison of
solutions to the equation of motion with (cosmological) observations.

The model as it stands is not complete since it does not
select a particular supersymmetric extension of the
standard model. It does provide a framework within which all
physical interactions can be accommodated and can be reduced to an underlying
topological structure, which is richer than this first investigation indicates.
Additional avenues to explore are the lattice structure of the anomalous
$U(1)$ symmetry, and its use as a horizontal symmetry for the mass and
mixing hierarchy[5]. In fact, the very occurrence of symmetry breaking on a
lattice of three-tori suggests that the junctions must
possess distinct senses of parity. A closer comparison with recent results on
black hole entropy is also warranted, since the solutions of the equation of
motion (5) should describe the black hole quantum states. In the boundary
condition $L=n\ell_{\rm Planck}$ on $L(T^3)$, the Schwarzschild radius in
Planck units now equals $n$, which fixes the black hole temperature and the
longest wavelength modes travelling along any of the six junctions of a
three-torus. According to the interpretation of (5), the black hole singularity
represents a region in space where all particle rest masses are of the order of
the Planck mass, and should be described mathematically by a soliton.

The presented results seem to lead to the notion that the combined
description of the standard model and gravity, is a ``unification without
unification''. That is, space-time topology (not fixed by general relativity)
leads to the standard model. Conversely, the Planck scale equation of motion
in the presence of strong gravitational effects requires no knowledge of
particle species and their interactions. Still, the link between them is
provided by the presence of both the (flat) $T^3$ and the (curved)
$S^1\times S^2$ prime manifold. Each represents a distinct branch of the
unification. Both hinge on the existence of a fundamental smallest scale.
It is in this respect that the topology of $Q$ and $L(T^3)$ provides a
conceptually different view of fundamental interactions.

\section*{Acknowledgements}
The author is indebted to G.~van Naeltwijck van Diosne, J.A.A.~Berendse-Vogels,
W.G.~Berendse, and M.A.R.~Bremer for valuable assistance.
This work was supported by NASA through Hubble Fellowship grant HF-01101.01-97A
awarded by the Space Telescope Science Institute, which is operated by the
Association of Universities for Research in Astronomy, Inc., for NASA under
contract NAS 5-26555.

\end{document}